\documentclass{jhep}

\title{NEGATIVE DIMENSIONAL INTEGRATION: "LAB TESTING" AT TWO LOOPS}
\author{Alfredo T. Suzuki and Alexandre G. M. Schmidt\\
Instituto de F\'{\i}sica Te\'orica -- Universidade Estadual Paulista,
R.Pamplona, 145, S\~ao Paulo SP, CEP 01405-900, Brazil\\
E-mail:\email{suzuki@power.ift.unesp.br, schmidt@power.ift.unesp.br }}

\abstract{
Negative dimensional integration method (NDIM) is a technique to deal with
$D$-dimensional Feynman loop integrals. Since most of the physical quantities
in perturbative Quantum Field Theory (pQFT) require the ability of
solving them, the quicker and easier the method to evaluate them the better.
The NDIM is a novel and promising technique, {\em ipso facto} requiring that
we put it to test in different contexts and situations and compare the
results it yields with those that we already know by other well-established
methods. It is in this perspective that we consider here the calculation 
of an on-shell two-loop three point function in a massless theory.
Surprisingly this approach provides twelve non-trivial results in terms
of double power series. More astonishing than this is the fact that we can
show these twelve solutions to be different representations for the same
well-known single result obtained via other methods. It really comes to us as
a surprise that the solution for the particular integral we are dealing with
is twelvefold degenerate. } 

\def\s{\sigma }
\begin{document}

\section{Introduction.}

Negative dimensional integration method (NDIM{}) was first devised by
Halliday \textit{et al }\cite{halliday,halliday2} with the aim of dealing
with Feynman loop integrals in quantum field theory (QFT). It combines the
powerful concepts of analytic continuation, translational invariance and 
dimensional regularization\cite{thooft,collins,wilson,nash} (DREG) in such a
way that the intricate positive dimensional integration is transformed into
negative dimensional integration of polynomial type. In practical terms,
what one needs to do is to solve systems of linear algebraic equations and
gaussian-like integrals.

Before plunging deeply in the mire of it, let us briefly make some comments.
First of all,  what do we mean by \textit{negative dimensions}? Obviously,
at their face value they can only be fictional. Nevertheless, in a stretch
of our imagination, if we allow ourselves just the possibility of their
existence, maybe we can bring into fruition something ever undreamed of
before. So, to begin with, let us mirror our reasoning with that behind
DREG. There, the all important concept is the parameter $D$, the space-time
dimension, allowed to assume complex values. However, this in no way means
that we want or will even define operations like scalar product\cite
{collins,wilson} in some general $D$ to have reality. The reason is quite
simple: Physics --- and even the common sense --- tells us that $D$ is a
positive integer number. We keep our eyes in this fact. One may think that
this way of expressing the result of a given Feynman integral in an
arbitrary space-time dimension is very elegant indeed, but at the end of the
day, one always has to look into the real physical world, that is, that of
positive integer $D$. Actually, in the whole process of DREG it is important
to keep $D$ arbitrary because the analyticity properties depend on the
space-time dimension and, of course,  any singularities do so depend on it
too. Wilson\cite{wilson} noted, very early, that the so-called integration
in $D$-dimensions is not in fact real literally. It only seems to be. It
only behaves like it is. 

So, we too define an integral in $D$-dimensions by means of an analytic
function and then work with general $D$, negative values included. We will
not attempt to make any sense of it or speculate about the $D<0$ "world".
What we do is just allow for it without trying to see how real --- or how
fictional, for that matter --- it is,  before doing the analytic{}
continuation{} back to $D>0$. In other words, we are not concerned with
seeking any meaning for a negative dimensional world nor seeking any new
Physics.

Let us sketch the methodology proper. The idea is quite simple: we
analytically continue the Feynman integral we want to evaluate into $D<0$
and solve it there. Then, the result we get there we bring it back into the
realm of $D>0$, by another analytic continuation.  As we shall see, the
whole procedure is much easier to do compared to positive dimensional
techniques. Up to now, for all the  Feynman integrals we have calculated
using NDIM{}, the results agree with the ones calculated in the positive
dimensional regime\cite{box,suzuki1,suzuki3}. This includes even some
light-cone gauge loop integrals\cite{suzuki2}, which knowingly are harder to
solve with other approaches.

The outline of our paper is as follows: in Sec.2 we solve explicitly a 
two-loop Feynman integral entering the two-loop radiative correction to the
massless triangle diagram using the negative--$D$ approach. Then, we  show
how to analytically continue the result to positive $D$ and give the result
for two on-shell external legs. In Sec.3 we conclude this work commenting on
some new results we have for massive one-loop and off-shell two-loop Feynman
integrals. Finally, in the Appendix we list all the other remaining degenerate
solutions, for the sake of completeness.

\section{On-Shell Two-loop Vertex.}

As we have mentioned earlier, our work here is done within the perpective of
checking NDIM methodology for $D$-dimensional Feynman integrals. Therefore,
we have chosen as our "lab test" for it, the evaluation of an integral
pertinent to the Feynman diagram of Fig.1. It is a two-loop graph with four
particles in the intermediate states, i.e., containing four propagators. For
simplicity we take the massless case and in order to compare our result with
the one already known in the literature, we take a particular limit of
two on-shell external legs. 

Since ours is a choice example, we cannot become too excited about NDIM and its
seemingly simplicity, yet one can convince himself that the task of solving
Feynman loop integrals is quite easy in this approach (at least in
principle). Of course, we can envisage and even anticipate some technical
difficulties in other contexts, which is inherent to the method, such as the
necessity of dealing with multi-indexed summations.

NDIM is implemented with a few simple steps: Firstly, we calculate gaussian
or gaussian-type integrals\footnote{in $D$-dimensional momentum space}, which
are not difficult to handle, and some books on QFT even list them in tables.
Then, one makes an expansion in Taylor series of the result obtained, and
compare it with the expansion in Taylor series of the original gaussian or
gaussian-type integral. Comparison term by term of both series then yield a
system of linear algebraic equations, with constraints arising from
intermediate multinomial expansions. NDIM{} requires that we
solve this system. The principle therefore is quite simple! Let us then take
a practical example and see how it works.

\FIGURE[r]{\ 
\begin{picture}(600,200)(000,0)
\thinlines
\put(250,200){\line(0,-1){50}}  %pra baixo (p)
\put(260,200){\vector(0,-1){30}}
\put(250,150){\line(-1,-1){90}} % diagonal a esquerda (t)
\put(150,70){\vector(1,1){20}}
\put(250,150){\line(1,-1){90}} % diagonal a direita (k)
\put(330,90){\vector(1,-1){20}}
\put(200,100){\line(1,0){100}} %fecha o triangulo
\put(250,100){\oval(100,70)[b]}
{\small 
\put(130,50){\makebox(0,0)[b]{$t=k-p$}} 
\put(360,50){\makebox(0,0)[b]{$k$}}
\put(230,200){\makebox(0,0)[b]{$p$}}
\put(200,120){\makebox(0,0)[b]{$q-p$}}
\put(300,120){\makebox(0,0)[b]{$q$}}
\put(250,50){\makebox(0,0)[b]{$q-r-k$}}
\put(250,105){\makebox(0,0)[b]{$r$}} }
\end{picture}
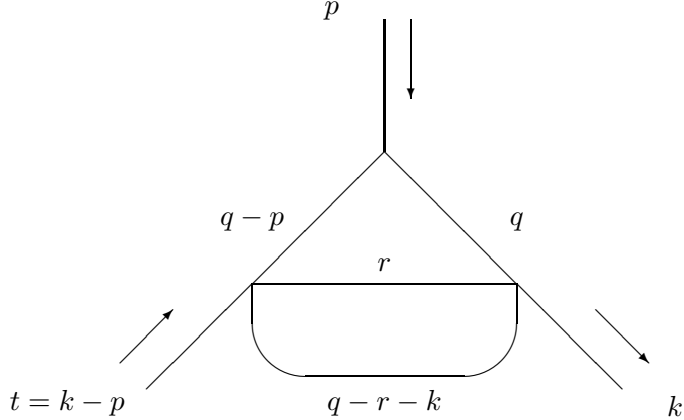
\caption{Two-loop three point vertex. } }

Let our launching-pad gaussian-like double integral be 

\begin{equation}
I = \int\int d^D\!r\; d^D\!q \;\;\exp\left[-\alpha q^2 -\beta (q-p)^2
-\gamma r^2 -\omega (q-r-k)^2\right] ,  \label{gauss}
\end{equation}

This clearly is a pertinent integral to the diagram of Fig.1. For reasons of
simplicity and future comparison, let us consider that two of the external
particles are real, i.e., let them be on-shell, namely, $k^2=t^2=0$.

Completing the square in the variable $q$ we can carry out the first
integration and get,

\begin{eqnarray}
I &=& \left(\frac{\pi}{\lambda}\right)^{\frac{1}{2} D} e^{-\beta
p^2+\frac{\beta^2p^2}{\lambda}+\frac{2\beta\omega pk}{\lambda}} \int
d^Dr\exp{\left(-\theta r^2-2\omega rk\right)} \exp{\left[\frac{1}{\lambda}  
\left(\omega^2r^2 \right.\right.}  \nonumber \\
&& \left.\left. +2\beta\omega pr+ 2\omega^2 rk\right)\right] ,
\end{eqnarray}
where $\theta=\gamma+\omega$ and $\lambda=\alpha+\beta+\omega$. Following
the same procedure we perform the remaining integration, remembering that
$t=k-p$ and that because of the on-shell condition $t^2=k^2=0$, $p^2=2pk$:

\begin{equation}
I = \left(\frac{\pi^2}{\phi}\right)^{\frac{1}{2} D} \exp{\left[ \frac{-1}{ 
\phi}(\alpha\beta\gamma+\alpha\beta\omega )p^2 \right]} ,
\end{equation}
where $\phi=\alpha\gamma+\alpha\omega+ \beta\gamma+ \beta\omega+ \gamma\omega
$. Expanding the exponential in Taylor series and using the multinomial
expansion in $\phi$, we get

\begin{eqnarray}
I &=&\pi^D \sum_{\{n_i=0\}}^\infty \frac{(-p^2)^{n_1+n_2} (-n_1-n_2-\frac{1}{ 
2} D)!} {n_1!n_2!n_3!n_4!n_5!n_6!n_7!}  \nonumber
\label{serie} \\
&&
\times\alpha^{n_1+n_2+n_3+n_4}\beta^{n_1+n_2+n_5+n_6}\gamma^{n_1+n_3+n_5+n_7}
\omega^{n_2+n_4+n_6+n_7} ,
\end{eqnarray}
where the summation indices must satisfy the constraint
$-n_1-n_2-\frac{1}{2}D=n_3+n_4+n_5+n_6+n_7$ which is the condition imposed by
the multinomial expansion.

Now comes the trick\cite{halliday,box,suzuki1,suzuki3,suzuki2} of negative
dimensions: Expand the integral (\ref{gauss}) in Taylor series,

\begin{eqnarray}
I &=&\sum_{i,j,l,m=0}^\infty \frac{(-1)^{i+j+l+m} \alpha^i
\beta^j\gamma^l\omega^m}{i!j!l!m!} \nonumber  \label{ndim} \\ 
&& \times\int\int d^Dq\;d^Dr\;(q^2)^i\left[(q-p)^2\right]^j (r^2)^l 
\left[(r-q+k)^2\right]^m ,
\end{eqnarray}
and let us define 
\begin{equation}  \label{Indim}
J_{NDIM} = \int d^Dq\int d^Dr \;(q^2)^i\left[(q-p)^2\right]^j
(r^2)^l\left[(r-q+k)^2\right]^m .
\end{equation}

We already note that this would be exactly the Feynman integral needed to be
evaluated for the diagram in Fig.1 if it were not for the positive exponents
$i,j,l$ and $m$. 

Comparing now (\ref{serie}) and (\ref{ndim}) we conclude that,

\begin{eqnarray}
J_{NDIM} &=& \frac{\pi^Dg(i,j,l,m)}{(-1)^{i+j+l+m}} \sum_{\{n_i=0\}}^\infty 
\frac{(-p^2)^{n_1+n_2}\Gamma(1-n_1-n_2-\frac{1}{2} D)}{ 
n_1!n_2!n_3!n_4!n_5!n_6!n_7!} \times  \nonumber \\
&& \!\!\!\delta_{n_1+n_2+n_3+n_4,i}\delta_{n_1+n_2+n_5+n_6,j}
\delta_{n_1+n_3+n_5+n_7,l}\delta_{n_2+n_4+n_6+n_7,m},
\end{eqnarray}
where 
\[
g(i,j,l,m)= \Gamma(1+i)\Gamma(1+j)\Gamma(1+l)\Gamma(1+m). 
\]

We can immediately see that there is a seven index summation\footnote{i.e.,
a "heptaple" series} and five equations linking them (remembering that one
comes from the constraint imposed by the multinomial expansion). Therefore,
altogether the system can be solved in twenty-one different ways with two
remaining series\footnote{the remnant double series}. Nine of them are
trivial solutions, which present no interest at all. The remaining twelve we
must solve one by one. In principle we do not know whether these are
equivalent or not\cite{box,boxnew,2loops}.  

A little bit of algebraic rearrangement yields, 
\begin{eqnarray}
J_{NDIM} &=& (-\pi)^D(p^2)^\sigma g(i,j,l,m)\Gamma(1-\sigma-\frac{1}{2} D)
\nonumber \\ &&\times \sum_{\{n_i=0\}}^\infty\frac{\delta_{n_1+n_2+n_3+n_4,i}
\delta_{n_1+n_2+n_5+n_6,j}\delta_{n_1+n_3+n_5+n_7,l}\delta_{n_2+n_4+n_6+n_7,m}}
{n_1!n_2!n_3!n_4!n_5!n_6!n_7!},
\end{eqnarray}
where we have defined $\sigma=i+j+l+m+D$.

It is a simple matter to write down a computer program that solves the
system exactly in all its twenty-one different ways. One of the solutions,
namely, that with remaining sum indices $n_2$ and $n_6$  is,

\begin{eqnarray}  \label{sol1}
S_1 &=& \frac{(-\pi)^D(p^2)^\sigma\Gamma(1+i)\Gamma(1+j)
\Gamma(1+l)\Gamma(1+m)\Gamma(1-\sigma-\frac{1}{2} D)}{\Gamma(1+\sigma) 
\Gamma(1-j-m-\frac{1}{2} D) \Gamma(1-l-\frac{1}{2} D)\Gamma(1+j-\sigma)} \\
&&\times \frac{1}{\Gamma(1+l+m+\frac{1}{2} D)} \sum_{n_2,n_6=0}^\infty \frac{ 
(-\sigma|n_2)(\frac{1}{2} D+l|n_2+n_6)(\sigma-j|n_6)}{n_2!n_6! (1-j-m-\frac{1 
}{2} D|n_2+n_6)} ,  \nonumber
\end{eqnarray}
where 
\[
(a|b)\equiv (a)_b = \frac{\Gamma(a+b)}{\Gamma(a)} 
\]
is the Pochhammer symbol\cite{bateman,grad,rainville} and we use one of its
properties, i.e., 
\begin{equation}  \label{prop}
(a|-k) = \frac{(-1)^k}{(1-a|k)} ,
\end{equation}
within the double series. Note that the sum above can be rewritten in terms
of hypergeometric functions $_2F_1$ if we use another property of the
Pochhammer symbol, i.e., $(a|b+c) = (a+b|c)(a|b)$. Now, a hypergeometric
function $_2F_1$ with unit argument can, within certain constraints in its
arguments, be summed\cite{bateman,grad,rainville},

\begin{equation}  \label{2f1}
_2F_1(a,b;c|1) = \frac{\Gamma(c)\Gamma(c-a-b)}{\Gamma(c-a)\Gamma(c-b)} .
\end{equation}

Rearranging first the $n_6$ sum we have,

\begin{eqnarray}
S_1 &=& \frac{(-\pi)^D(p^2)^\sigma\Gamma(1+i)\Gamma(1+j)
\Gamma(1+l)\Gamma(1+m)\Gamma(1-\sigma-\frac{1}{2} D)}{\Gamma(1+\sigma) 
\Gamma(1-j-m-\frac{1}{2} D) \Gamma(1-l-\frac{1}{2} D)\Gamma(1+j-\sigma)} 
\sum_{n_2=0}^\infty  \nonumber \\
&&\times \frac{(-\sigma|n_2)(\frac{1}{2} D+l|n_2)}{n_2!(1-j-m-\frac{1}{2}
D|n_2)} \sum_{n_6=0}^\infty \frac{(\sigma-j|n_6)(\frac{1}{2} D+l+n_2|n_6)}{ 
n_6! (1-j-m-\frac{1}{2} D +n_2|n_6)}  \nonumber \\
&&\times \frac{1}{\Gamma(1+l+m+\frac{1}{2} D)} ,
\end{eqnarray}
where the second series is by definition the gaussian hypergeometric
function, $ _2F_1$. Using (\ref{2f1}) we can sum it and then sum also the
series in $n_2$, to get,

\begin{eqnarray}  \label{result}
S_1 &=& \frac{(-\pi)^D(p^2)^\sigma\Gamma(1+i)\Gamma(1+j)
\Gamma(1+l)\Gamma(1+m)\Gamma(1-\sigma-\frac{1}{2} D)}{\Gamma(1+\sigma)
\Gamma(1-l-\frac{1}{2} D)\Gamma(1+j-\sigma)\Gamma(1+l+m+\frac{1}{2} D)} 
\nonumber \\
&& \times\frac{\Gamma(1-l-m-D)}{\Gamma(1+i-\sigma)\Gamma(1-m-\frac{1}{2} D)}.
\end{eqnarray}

Now, grouping the gamma functions in convenient Pochhammer symbols and using
(\ref{prop}) we analytic continue to negative values of the exponents
$i,j,l,$ and $m$ and we go back to positive $D$ to get,

\begin{eqnarray} \label{geral}
S_1^{AC} &=& \pi^D(p^2)^\sigma(-i|\sigma)(-j|\sigma)(-l|-m-\frac{1}{2}
D)(-m|\sigma-i-j-\frac{1}{2} D)  \nonumber \\
&& \times (\sigma+\frac{1}{2} D|-2\sigma-\frac{1}{2} D)(D+l+m|-l-\frac{1}{2}
D).
\end{eqnarray}

This is the general result, in Euclidean space, for the Feynman graph of
Fig.1.

In the important particular case when $i=j=l=m=-1$, the result is

\begin{equation}  \label{onshell}
S_1^{AC} = \frac{\pi^D(p^2)^{D-4}\Gamma^2(D-3)\Gamma^2(\frac{1}{2} D-1)
\Gamma(2-\frac{1}{2} D)\Gamma(4-D)}{\Gamma(D-2)\Gamma(\frac{3}{2} D-4)},
\end{equation}
which is the well-known result in $D$-dimensions\cite{hath}.

We ask immediately: What is the result that the other solutions provide? The
answer is as surprising as it could be: the result is the same. All the
twelve non-trivial solutions, even distinct, give the same result. It is an
amazing feature revealed by NDIM. 

Just to check on this, let us consider another solution, for instance, the
one where $n_2$ and $n_5$ are the remaining sum indices, 

\begin{eqnarray}
S_2 &=& \frac{(-\pi)^D(p^2)^\sigma \Gamma(1+i)\Gamma(1+j) \Gamma(1+l)\Gamma(1+m)
\Gamma(1-\sigma-\frac{1}{2} D)} {\Gamma(1+\sigma)\Gamma(1-\sigma-m-\frac{1}{2 
} D)\Gamma(1+j-\sigma)\Gamma(1+l+m+\frac{1}{2} D) }  \nonumber \\
&&\times \frac{1}{\Gamma(1+i+m+\frac{1}{2} D)} \sum_{n_2,n_5=0}^\infty \frac{ 
(-1)^{n_2+n_5}(-\sigma|n_2)(-j+\sigma|n_5)}{n_2!n_5!(1-\sigma-m-\frac{1}{2}
D|n_2-n_5) }  \nonumber \\
&&\times \frac{1}{(1+i+m+\frac{1}{2} D|n_5-n_2)}.
\end{eqnarray}

As in the previous case we can sum both series. Using (\ref{prop}) one
can rewrite the $n_5$ series and identify it as a $_2F_1$ function, 
\begin{eqnarray}
S_2 &=& (-\pi)^D(p^2)^\sigma P_2(i,j,l,m;D)\sum_{n_2=0}^\infty \frac{ 
(-1)^{n_2}(-\sigma|n_2)}{n_2!(1-\sigma-m-\frac{1}{2} D|n_2)} \\
&&\times\frac{1}{(1+i+m+\frac{1}{2} D|-n_2)} \sum_{n_5=0}^\infty \frac{ 
(\sigma-j|n_5)(\sigma+m+\frac{1}{2} D-n_2|n_5)}{n_5!(1+i+m+\frac{1}{2}
D-n_2|n_5)},  \nonumber
\end{eqnarray}
where 
\begin{eqnarray}
P_2(i,j,l,m;D) &=& \frac{\Gamma(1+i)\Gamma(1+j) \Gamma(1+l)\Gamma(1+m)
\Gamma(1-\sigma-\frac{1}{2} D)} {\Gamma(1+\sigma)\Gamma(1-\sigma-m-\frac{1}{2 
} D)\Gamma(1+j-\sigma)}  \nonumber \\
&&\times\frac{1}{\Gamma(1+l+m+\frac{1}{2} D)\Gamma(1+i+m+\frac{1}{2} D)}, 
\nonumber
\end{eqnarray}

Summing the $n_5$ series with formula (\ref{2f1}) and using again
(\ref{prop}) to rewrite the $n_2$ series, we get 

\begin{eqnarray}
S_2 &=& (-\pi)^D(p^2)^\sigma P_2(i,j,l,m;D)\frac{\Gamma(1+i+m+\frac{1}{2}
D)\Gamma(1+i+j-2\sigma)}{\Gamma(1+i-\sigma)\Gamma(1-l-\frac{1}{2} D)} 
\nonumber \\
&&\times \sum_{n_2=0}^\infty \frac{(-\sigma|n_2)(\frac{1}{2} D+l|n_2)}{ 
n_2!(1-\sigma-m-\frac{1}{2} D|n_2)}.
\end{eqnarray}

This series is by definition a summable $_2F_1$ function; with the help of
eq.(\ref {2f1}), we get the expression (\ref{result}) which leads to the
correct result (\ref{onshell}).

The reader can prove, following the same procedure, that all the twelve
solutions provide the correct result. The question that arises is: Why is
this so? We have no answer to this puzzle at the moment and can only
conjecture that maybe if the remaining series has unity argument and is
summable then the result will be degenerate. Of course, further research is
necessary in order to prove or disprove this conjecture.

\section{Conclusion.}

Our two-loop "lab testing" for NDIM approach to calculate Feynman integrals
has revealed some very interesting features of the method. The methodology is
quite simple: solving gaussian integrals and systems of linear algebraic
equations. NDIM yielded twelve non-trivial solutions which give the same
result, eq.(\ref{geral}), for the general case, i.e.,  $D$-dimensions and
arbitrary exponents of 
propagators. This work encourages us to tackle a more difficult task: the
calculation of massive four point one-loop integrals\cite{box,boxnew} and
off-shell two-loop Feynman graphs  \cite{2loops}. Work in this line is in
progress, and we have already obtained some more encouraging results. For the
latter, for example, a new surprise with  NDIM yielding twenty-four
distinct (and new!) results, some of them in terms of Appel's\cite{appel}
hypergeometric functions $F_4$ which are simpler than the usual
dilogarithms\cite{letb}. These new results will be the subject addressed in
our shortly forthcoming paper. 

\acknowledgments{\ AGMS wishes to thank CNPq (Conselho Nacional de
Desenvolvimento Cient\'{\i}fico e Tecnol\'ogico, Brasil) for the financial
support. }

\appendix

\section{The Remaining Solutions.}

For the sake of completeness, in this appendix we list the remaining ten
non-trivial solutions since in the article proper we have explicitly shown
only two of them, and that these two give the correct result. All 
of these can be summed and analytically continued to positive $D$ with the
same ideas we used in section 2.

\begin{eqnarray}  \label{sol3}
S_3 &=& \frac{(-\pi)^D(p^2)^\s\Gamma(1+i)\Gamma(1+j)
\Gamma(1+l)\Gamma(1+m)\Gamma(1-\sigma-\frac{1}{2} D)}{\Gamma(1+\sigma+l+%
\frac{1}{2} D)\Gamma(1-l-\frac{1}{2} D) \Gamma(1+i-\sigma)\Gamma(1+j-\sigma)}
\\
&&\times \frac{1}{\Gamma(1+l+m+\frac{1}{2} D)} \sum_{n_4,n_6=0}^\infty \frac{%
(-j+\sigma|n_6)(\frac{1}{2} D+l|n_4+n_6)(-i+\sigma|n_4)}{n_4!n_6! (1+l+\frac{%
1}{2} D+\sigma|n_4+n_6)} ,  \nonumber
\end{eqnarray}

\begin{eqnarray}  \label{sol4}
S_4 &=& \frac{(-\pi)^D(p^2)^\s\Gamma(1+i)\Gamma(1+j)
\Gamma(1+l)\Gamma(1+m)\Gamma(1-\sigma-\frac{1}{2} D)}{\Gamma(1+j+l+\frac{1}{2%
} D)\Gamma(1+i+m+\frac{1}{2} D)
\Gamma(1+j-\sigma)\Gamma(1+l+m+\frac{1}{2} D)}\\
&&\times \frac{1}{\Gamma(1+i-\sigma)} \sum_{n_4,n_5=0}^\infty
\frac{(-1)^{n_4+n_5}(-i+\sigma|n_4)(-j+\sigma|n_5)}{n_4!n_5!
(1+j+l+\frac{1}{2} D|n_4-n_5)(1+i+m+\frac{1}{2} D|n_5-n_4)}
,\nonumber\end{eqnarray} 

\begin{eqnarray}  \label{sol5}
S_5 &=& \frac{(-\pi)^D(p^2)^\s\Gamma(1+i)\Gamma(1+j)
\Gamma(1+l)\Gamma(1+m)\Gamma(1-\sigma-\frac{1}{2} D)}{\Gamma(1+i+l+\frac{1}{2%
} D)\Gamma(1+j+m+\frac{1}{2} D)
\Gamma(1+j-\sigma)\Gamma(1+l+m+\frac{1}{2} D)}\\
&&\times \frac{1}{\Gamma(1+i-\sigma)}\sum_{n_3,n_6=0}^\infty
\frac{(-1)^{n_3+n_6}(-i+\sigma|n_3)(-j+\sigma|n_6)}{n_3!n_6!
(1+i+l+\frac{1}{2} D|n_6-n_3)(1+j+m+\frac{1}{2} D|n_3-n_6)}
,\nonumber\end{eqnarray} 

\begin{eqnarray}  \label{sol6}
S_6 &=& \frac{(-\pi)^D(p^2)^\s\Gamma(1+i)\Gamma(1+j)
\Gamma(1+l)\Gamma(1+m)\Gamma(1-\sigma-\frac{1}{2} D)}{\Gamma(1+\sigma+m+%
\frac{1}{2} D)\Gamma(1-m-\frac{1}{2} D) \Gamma(1+i-\sigma)\Gamma(1+j-\sigma)}
\\
&&\times \frac{1}{\Gamma(1+l+m+\frac{1}{2} D)} \sum_{n_3,n_5=0}^\infty \frac{%
(-i+\sigma|n_3)(-j+\sigma|n_5)(\frac{1}{2} D+m|n_3+n_5)}{n_3!n_5! (1+m+\frac{%
1}{2} D+\sigma|n_3+n_5)} ,  \nonumber
\end{eqnarray}

\begin{eqnarray}  \label{sol7}
S_7 &=& \frac{(-\pi)^D(p^2)^\s\Gamma(1+i)\Gamma(1+j)
\Gamma(1+l)\Gamma(1+m)\Gamma(1-\sigma-\frac{1}{2} D)}{\Gamma(1+\sigma)%
\Gamma(1-i-m-\frac{1}{2} D) \Gamma(1+i-\sigma)\Gamma(1-l-\frac{1}{2} D)} \\
&&\times \frac{1}{\Gamma(1+l+m+\frac{1}{2} D)} \sum_{n_2,n_4=0}^\infty \frac{%
(-\sigma|n_2)(-i+\sigma|n_4)(\frac{1}{2} D+l|n_2+n_4)}{n_2!n_4! (1-i-m-\frac{%
1}{2} D|n_2+n_4)} ,  \nonumber
\end{eqnarray}

\begin{eqnarray}  \label{sol8}
S_8 &=& \frac{(-\pi)^D(p^2)^\s\Gamma(1+i)\Gamma(1+j)
\Gamma(1+l)\Gamma(1+m)\Gamma(1-\sigma-\frac{1}{2} D)}{\Gamma(1+\sigma)%
\Gamma(1+j+m+\frac{1}{2} D) \Gamma(1+i-\sigma)\Gamma(1-m-\frac{1}{2}
D-\sigma)\Gamma(1+l+m+\frac{1}{2} D)}  \nonumber \\
&&\times \sum_{n_2,n_3=0}^\infty \frac{(-1)^{n_2+n_3}(-\sigma|n_2)
(-i+\sigma|n_3)}{n_2!n_3! (1+j+m+\frac{1}{2} D|n_3-n_2) (1-m-\frac{1}{2}
D-\sigma|n_2-n_3)} ,
\end{eqnarray}

\begin{eqnarray}  \label{sol9}
S_9 &=& \frac{(-\pi)^D(p^2)^\s\Gamma(1+i)\Gamma(1+j)
\Gamma(1+l)\Gamma(1+m)\Gamma(1-\sigma-\frac{1}{2} D)}{\Gamma(1+\sigma)%
\Gamma(1+i+l+\frac{1}{2} D) \Gamma(1+j-\sigma)\Gamma(1-l-\frac{1}{2}
D-\sigma)\Gamma(1+l+m+\frac{1}{2} D)}  \nonumber \\
&&\times \sum_{n_1,n_6=0}^\infty \frac{(-1)^{n_1+n_6}(-\sigma|n_1)
(-j+\sigma|n_6)}{n_1!n_6! (1+i+l+\frac{1}{2} D|n_6-n_1) (1-l-\frac{1}{2}
D-\sigma|n_1-n_6)} ,
\end{eqnarray}

\begin{eqnarray}  \label{sol10}
S_{10} &=& \frac{(-\pi)^D(p^2)^\s\Gamma(1+i)\Gamma(1+j)
\Gamma(1+l)\Gamma(1+m)\Gamma(1-\sigma-\frac{1}{2} D)}{\Gamma(1+\sigma)%
\Gamma(1-m-\frac{1}{2} D) \Gamma(1+j-\sigma)\Gamma(1-j-l-\frac{1}{2} D)} \\
&&\times \frac{1}{\Gamma(1+l+m+\frac{1}{2} D)} \sum_{n_1,n_5=0}^\infty \frac{%
(-\sigma|n_1)(-j+\sigma|n_5)(\frac{1}{2} D+m|n_1+n_5)}{n_1!n_5! (1-j-l-\frac{%
1}{2} D|n_1+n_5)} ,  \nonumber
\end{eqnarray}

\begin{eqnarray}  \label{sol11}
S_{11} &=& \frac{(-\pi)^D(p^2)^\s\Gamma(1+i)\Gamma(1+j)
\Gamma(1+l)\Gamma(1+m)\Gamma(1-\sigma-\frac{1}{2} D)}{\Gamma(1+\sigma)%
\Gamma(1+j+l+\frac{1}{2} D) \Gamma(1+i-\sigma)\Gamma(1-l-\frac{1}{2}
D-\sigma)\Gamma(1+l+m+\frac{1}{2} D)}  \nonumber \\
&&\times \sum_{n_1,n_4=0}^\infty \frac{(-1)^{n_1+n_4}(-\sigma|n_1)
(-i+\sigma|n_4)}{n_1!n_4! (1+j+l+\frac{1}{2} D|n_4-n_1) (1-l-\frac{1}{2}
D-\sigma|n_1-n_4)} ,
\end{eqnarray}

\begin{eqnarray}  \label{sol12}
S_{12} &=& \frac{(-\pi)^D(p^2)^\s\Gamma(1+i)\Gamma(1+j)
\Gamma(1+l)\Gamma(1+m)\Gamma(1-\sigma-\frac{1}{2} D)}{\Gamma(1+\sigma)%
\Gamma(1-i-l-\frac{1}{2} D) \Gamma(1+i-\sigma)\Gamma(1-m-\frac{1}{2} D)} \\
&&\times \frac{1}{\Gamma(1+l+m+\frac{1}{2} D)} \sum_{n_1,n_2=0}^\infty \frac{%
(-\sigma|n_1)(-i+\sigma|n_3)(\frac{1}{2} D+m|n_1+n_3)}{n_1!n_2! (1-i-l-\frac{%
1}{2} D|n_1+n_3)} ,  \nonumber
\end{eqnarray}


\begin{thebibliography}{99}
\bibitem{halliday}  I.G.Halliday, R.M.Ricotta, Phys.Lett.\textbf{B193},
2(1987)241; R.M.Ricotta, \textit{Topics in Field Theory}, (Ph.D. Thesis,
Imperial College, 1987).

\bibitem{halliday2}  G.V.Dunne, I.G.Halliday, Phys.Lett.\textbf{B193} 
,2(1987)247.

\bibitem{thooft}  G.'t Hooft, M.Veltman, Nucl.Phys.\textbf{B44} (1972)189;
C.G. Bollini, J.J. Giambiagi, Nuovo Cim. \textbf{B12} (1972)20.

\bibitem{collins}  J.C.Collins, \textit{Renormalization} (Cambridge
Univ.Press, 1984).

\bibitem{wilson}  K.G.Wilson, Phys.Rev.\textbf{D7}, 10 (1973) 2911. See the
very clear discussion of $D$-dimensional integration in the appendix.

\bibitem{nash}  C.Nash,\textit{Relativistic Quantum Fields} (Academic Press,
1978).

\bibitem{box}  A.T.Suzuki,A.G.M.Schmidt, submitted to Nucl.Phys.{\bf B}
(1997);hep-th/9707187. 

\bibitem{suzuki1}  {A.T.Suzuki,R.Ricotta, \textit{Topics on Theoretical
Physics - Festschrift for P.L.Ferreira}, (1995)219, V.C.Aguilera-Navarro 
\textit{el al} (Ed.) }

\bibitem{suzuki3}  I.G.Halliday, R.M.Ricotta, A.T.Suzuki, \textit{XVII
Brazilian Meeting on Particles and Fields}, (1996)495, A.J. da Silva \textit{ 
et al} (Ed.).

\bibitem{suzuki2}  {A.T.Suzuki, R.M.Ricotta, \textit{XVI Brazilian Meeting
on Particles and Fields}, (1995) 386, C.O. Escobar (Ed.). }

\bibitem{boxnew}  A.T.Suzuki, A.G.M.Schmidt, in preparation (1997).

\bibitem{2loops}  A.T.Suzuki, A.G.M.Schmidt, in preparation (1997).

\bibitem{bateman}  A.Erd\'elyi, W.Magnus, F.Oberhettinger and F.Tricomi, 
\textit{Higher Transcendental Functions} (McGraw-Hill, 1953).

\bibitem{grad}  I.S.Gradstein, I.M.Rhizik, \textit{Table of Integrals,
Series and Products} (Academic Press, 1994).

\bibitem{rainville}  E.D.Rainville, \textit{Special Functions} (Chelsea
Pub.Co., 1960).

\bibitem{hath}  S.J.Hathrell, Ann.Phys.\textbf{139} (1982)136.

\bibitem{appel}  P.Appel, J. Kamp\'e de Feriet, \textit{Fonctions
Hyperg\'eom\'etriques et Hypersph\'eriques. Polynomes D'Hermite}
(Gauthiers-Villars, Paris 1926).

\bibitem{letb}  N.I.Ussyukina, A.I.Davydychev, Phys.Lett.\textbf{B332} 
(1994)159.
\end{thebibliography}
\end{document}